\documentclass[pre,aps,twocolumn,superscriptaddress]{revtex4-1}
\pdfoutput=1
\usepackage{amssymb}
\usepackage{epsfig}
\usepackage{amsmath}
\usepackage{subfigure}
\usepackage{graphicx}
\usepackage{textcomp}
\usepackage{url}
\usepackage{float}
\usepackage{color}
\usepackage{cases}
\usepackage[normalem]{ulem}

\usepackage[colorlinks=true, urlcolor=blue, anchorcolor=blue, citecolor=blue,filecolor=blue,linkcolor=blue,menucolor=blue%,pagecolor=blue
]{hyperref}

\usepackage{color}

\begin{document}
\title{Fluctuations of a swarm of Brownian bees}

\author{Maor Siboni}
\email{maor.siboni@mail.huji.ac.il}
\affiliation{Racah Institute of Physics, Hebrew University of
Jerusalem, Jerusalem 91904, Israel}
\author{Pavel Sasorov}
\email{pavel.sasorov@gmail.com}
\affiliation{Institute of Physics CAS, ELI Beamlines, 182 21 Prague, Czech Republic}
\affiliation{Keldysh Institute of Applied Mathematics, Moscow 125047, Russia}
\author{Baruch Meerson}
\email{meerson@mail.huji.ac.il}
\affiliation{Racah Institute of Physics, Hebrew University of
Jerusalem, Jerusalem 91904, Israel}

\begin{abstract}
The ``Brownian bees" model describes
an ensemble of $N$ independent branching Brownian particles. When a particle branches into two particles, the particle farthest
from the origin is eliminated so as to keep the number of particles constant. In the limit of $N\to \infty$,  the spatial density of the particles is governed by
the solution of a free boundary problem  for a reaction-diffusion equation. At long times the particle density approaches a  spherically symmetric steady state solution with a compact support. Here we study fluctuations of the ``swarm of bees" due to the random character of the branching Brownian motion in the limit of large but finite $N$. We consider a one-dimensional setting and focus on two fluctuating quantities: the swarm center of mass $X(t)$ and the swarm radius $\ell(t)$. Linearizing a pertinent  Langevin equation around the deterministic steady state solution, we calculate the two-time covariances of $X(t)$ and $\ell(t)$. The variance of $X(t)$ directly follows from the covariance of $X(t)$, and it scales as $1/N$ as to be expected from the law of large numbers. The variance of $\ell(t)$ behaves differently: it exhibits an anomalous scaling $(1/N) \ln N$. This anomaly appears because all spatial scales, including a narrow region near the edges of the swarm where only a few particles are present, give a significant contribution to the variance. We argue that the variance of $\ell(t)$ can be obtained from the covariance of $\ell(t)$ by introducing a cutoff at the microscopic time $1/N$ where the continuum Langevin description breaks down. Our theoretical predictions are in good agreement with Monte-Carlo simulations of the microscopic model. Generalizations to higher dimensions are briefly discussed.

\end{abstract}

\maketitle
\nopagebreak

\section{Introduction}
\label{intSec}

This work touches two distinct basic concepts of nonequilibrium statistical physics: branching Brownian motion (BBM) and nonequilibrium steady states (NESSs). BBM -- a combination of Brownian motion of particles  and their random branching -- was extensively studied in the past \cite{McKean,Bramson}, and it continues to attract a great deal of attention from physicists \cite{BD2009,Mueller,Ramola1,DMS} and mathematicians.

In its turn, NESSs of ensembles of reacting and diffusing particles occupy an important niche of nonequilibrium statistical mechanics \cite{JL1993,JL2004,Bodineau2010,Hurtado2013,M2015}. The NESSs capture in a simple way different aspects of  physics of more complicated dissipative systems, both living and non-living.

Here we study a simple but nontrivial example of a NESS in a particle-conserving version of branching Brownian motion (BBM), which is known under the name of ``Brownian bees" \cite{bees1,bees2,MS2021}. The model involves $N$ independent particles (bees) located in a $d$-dimensional space.  On a small
time interval $\Delta t$ each particle branches into two particles with probability $\Delta t$. It also performs,
with the  complementary probability $1-\Delta t$,  continuous-time Brownian motion with diffusion constant $1$.   When a branching event occurs, the particle which is farthest from
the origin is immediately removed, so as to keep the number of particles constant at all times.
The NESS
of this relatively simple system has some interesting and non-trivial properties that will be the focus of this paper.

An additional motivation to study the Brownian bees model is that it is a close relative  of a whole family of Brunet-Derrida $N$-particle models: branching Brownian models with selection, the studies of which were initiated in Refs. \cite{BDMM2006,BDMM2007}. A Brunet-Derrida $N$-particle model
consists of a population of $N$ particles which undergo BBM. When a branching event occurs, the particle of the lowest fitness is removed. The models differ among themselves  by their \emph{fitness function} -- thus mimicking different aspects of  biological selection -- and by the dimension of space. The
original papers \cite{BDMM2006,BDMM2007} considered discrete-time processes, but much of
the subsequent work \cite{BG2010,DR2011,BBS2013,M2016,BZ2018,DMFPSL}, including the studies of the Brownian bees model  \cite{bees1,bees2,MS2021} and its barocentric variant  \cite{ABLT,barocentric}, has focused on the continuous-time branching Brownian motion.

Here we consider the Brownian bees model on the line, $d=1$. It has been rigorously shown in Ref. \cite{bees1} that, in the limit of $N\to \infty$, the coarse-grained spatial density $u(x,t)\geq 0$ of the particles, normalized by $N$, is governed by a deterministic free boundary problem:
\begin{eqnarray}
  &&\partial_t u (x,t) = \partial_x^2 u(x,t)+u(x,t)\,,\quad |x|\leq \ell(t)\,, \label{INT10}\\
  &&u(x,t)=0\,, \quad |x|> \ell(t)\,,
  \label{LAN110} \\
  &&\int_{-\ell(t)}^{\ell(t)}u(x,t)\, dx =1\,.
  \label{LAN120}
\end{eqnarray}
$u(x,t)$ is continuous at $|x|=\ell(t)$, and an initial condition should be specified. As one can see, the compact support of $u(x,t)$, at all $t>0$, is centered at the origin. Effectively, there are two absorbing walls, at $x=\pm \ell(t)$, which move in synchrony so as to keep the number of particles constant at all times.

Furthermore, it has been proven in Ref.~\cite{bees2} that, at long times, the solution of the deterministic problem (\ref{INT10})-(\ref{LAN120}) approaches a unique steady state
\begin{equation}\label{Ux}
U(x) =\left\{
\begin{array}{lr}{\displaystyle \frac{1}{2}}\cos x\,, & ~~~|x| \le \ell_0,  \\
\\
0\,,& ~~~|x|>\ell_0,
\end{array}
\right.
\end{equation}
%\begin{numcases}
%{{U(x)} =}\frac{1}{2}\cos x\,, & $|x| \le \ell_0$, \label{Ux} \\
%0\,,& $|x|>\ell_0$, \label{outsidex}
%\end{numcases}
where $\ell_0=\pi/2$. Correspondingly, $\ell(t)$ approaches $\ell_0$.

A natural next step is to study \emph{fluctuations} in the Brownian bees model  at large but finite $N$, caused by the random character of the elemental processes of the branching Brownian motion. \emph{Persistent} fluctuations, including large deviations, of the swarm size $\ell(t)$ have been recently considered in Ref. \cite{MS2021}.  Here we study typical (that is small) Gaussian fluctuations of the swarm center of mass $X(t)$ and the swarm radius $\ell(t)$. Our goals in this work are to calculate the steady-state two-time covariances of $X(t)$ and $\ell(t)$, which we call $g_X(\tau)$ and $g_{\ell} (\tau)$, respectively. Here $\tau$ is the time between two measurements of the corresponding quantities in the steady state. To this end we introduce a pertinent Langevin equation for the Brownian bees model and linearize it around the deterministic steady state solution~(\ref{Ux}). %and (\ref{outsidex}).
Solving the linearized equation, we obtain explicit expressions for small random deviations of $X(t)$ and $\ell(t)$ from their deterministic stationary values $0$ and $\ell_0$, respectively.
Then we use these solutions to compute the functions $g_X(\tau)$ and $g_{\ell} (\tau)$.

We find that the covariance $g_X(\tau)$ remains finite at $\tau \to 0$ and yields the variance of $X(t)$ which scales as $1/N$, as to be expected from the law of large numbers.

The covariance $g_{\ell}(\tau)$ behaves markedly differently. At $\tau\ll 1$ it exhibits a logarithmic scaling with $\tau$, giving an unexpected example of a system with $1/f$ noise.  Furthermore, $g_{\ell}(\tau)$  formally diverges logarithmically as $\tau \to 0$.  As we argue, the continuum Langevin description of the Brownian bees breaks down when $\tau$ becomes comparable with the microscopic time of the model, $1/N$. Using the microscopic time as a cutoff, we evaluate, with a logarithmic in $N$ accuracy, the variance which scales as $(1/N) \ln N$. This and other theoretical predictions are in good agreement with Monte Carlo simulations of the microscopic model that we performed.

Here is a plan of the remainder of this paper. In Sec.~\ref{lanSec} we introduce a coarse-grained stochastic formulation of the problem, based on the Langevin equation for the BBM. In Sec.~\ref{linSec} we formulate a linearized version of the problem by expanding the solution around the (deterministic) steady-state solution (\ref{Ux}). %and (\ref{outsidex}).
In the same Sec.~\ref{linSec} we obtain the  solution to the linearized problem at long times. This solution is then used in Secs. \ref{comSec} and \ref{lSec}
to calculate the two-time covariances, and then the variances, of $X(t)$ and $\ell(t)$, respectively. The theoretical predictions are compared with results of Monte-Carlo simulations.  We summarize and briefly discuss our results in Sec. \ref{sumSec}. The Appendix includes a brief description of the Monte-Carlo simulation algorithm.

\section{Langevin equation}
\label{lanSec}

To account for fluctuations resulting from
the intrinsic randomness of the elemental processes of the BBM, one must go beyond the deterministic equation~(\ref{INT10}). At large $N$ typical fluctuations are small, and they are captured by the Langevin equation: a stochastic partial differential equation \cite{vK,Gardiner}. In our case the Langevin equation can be written as
\begin{equation}\label{LAN70}
\partial_t u (x,t) = \partial_x^2 u(x,t)+u(x,t)+\tilde{R}(u,x,t),\;\; |x|\leq \ell(t).
\end{equation}
This equation replaces the mean-field equation~(\ref{INT10}), whereas Eqs.~(\ref{LAN110}) and
(\ref{LAN120}) remain unchanged. Now the solution $u(x,t)$ and the size of its support $\ell(t)$ are random functions of their variables. The multiplicative noise term $\tilde{R}(u,x,t)$ comes from two independent multiplicative noises:  the branching noise $\tilde{R}_b(u,x,t)$ and the noise of Brownian motion $\tilde{R}_d(u,x,t)$:
\begin{eqnarray}
%\label{LAN80}
&&\tilde{R}(u,x,t)=\tilde{R}_b(u,x,t)+\tilde{R}_d(u,x,t)\,,\nonumber\\
\label{LAN90}
&&\tilde{R}_b(u,x,t)=\frac{\sqrt{u}}{\sqrt{N}}\eta(x,t)\,,\\
\label{LAN100}
&&\tilde{R}_d(u,x,t)=\frac{1}{\sqrt{N}}\partial_x\left[\chi(x,t)\sqrt{2u}\right]\,,
\end{eqnarray}
where $\eta(x,t)$ and $\chi(x,t)$ are two independent Gaussian white noises with zero mean:
\begin{eqnarray}
  \left\langle\eta(x_1,t_1)\eta(x_2,t_2)\right\rangle &=& \left\langle\chi(x_1,t_1)\chi(x_2,t_2)\right\rangle \nonumber \\
  &=& \delta(x_1-x_2)\delta(t_1-t_2) \,. \label{LAN140}
\end{eqnarray}

A few words are in order about the origin of $R_b$ and $R_d$. The branching noise
$R_b$ can be derived from the exact master equation for the branching process. For typical fluctuations in a system of many particles one can use the van Kampen system size expansion to approximate the master equation by a Fokker-Planck equation, see \textit{e.g.} Ref. \cite{vK,Gardiner}. In the equivalent language of the Langevin equation one obtains Eq.~(\ref{LAN90}).  The Brownian motion noise term $R_d$ is best known
in the context of a large-scale and long-time description of the dynamics of a lattice gas of independent random walkers, see \textit{e.g.} Ref. \cite{Spohn}.

Differentiating the conservation law~(\ref{LAN120}) with respect to time and using Eq.~(\ref{LAN70}), we obtain
\begin{equation}\label{LAN150}
\partial_x u[-\ell(t),t]-\partial_x u[\ell(t),t] =1+\int\limits_{-\ell(t)}^{\ell(t)}\tilde{R}\, dx\, .
\end{equation}
We will use this equation instead of Eq.~(\ref{LAN120}) at $t>t_0$, where $t_0$ is the initial time.

One should bear in mind that a Langevin description provides a \emph{macroscopic} description of fluctuations in microscopic models like the one we are dealing with here. In this particular case it is accurate only at times much longer than the typical time between two consecutive branching events: $\Delta t \gg 1/N$, and at distances much larger than the typical interparticle distance. The latter is of order $1/N$ in the bulk of the swarm, but much larger -- of order
$1/\sqrt{N}$ -- at the edges of the swarm, where the macroscopic density, see Eq.~(\ref{Ux}), %and (\ref{outsidex}),
approaches zero.

\section{Linearization and solution}
\label{linSec}
\subsection{Linearization}
\label{linSubSec}

The small parameter $1/\sqrt{N} \ll 1$  in the noise term in Eq.~(\ref{LAN70})  calls for a perturbation expansion. In the leading order  one can simply linearize Eq.~(\ref{LAN70}) around the deterministic steady state~(\ref{Ux}). %and (\ref{outsidex}).
We write
\begin{eqnarray}
\label{LIN10}
&&u(x,t)=U(x) + v(x,t)\,,\quad |v|\ll 1\,;\quad v\propto \frac{1}{\sqrt{N}} \,,\\
\label{LIN20}
&&\ell(t)=\frac{\pi}{2} + \delta\ell(t)\,,\quad |\delta\ell|\ll 1\,; \quad \delta\ell \propto \frac{1}{\sqrt{N}}.
\end{eqnarray}
The linearized versions of Eqs.~(\ref{LAN70}), (\ref{LAN110}) and~(\ref{LAN150}) are
\begin{eqnarray}
\label{LIN30}
&&\partial_t v(x,t)- \partial_x^2 v(x,t) -v(x,t) =\tilde{R}(x,t)\,,\\
\label{LIN40}
&&-\frac{1}{2}\delta\ell(t)+v\left(\pm\frac{\pi}{2},t\right)=0\,,\\
\label{LIN50}
&&\partial_x v\left(-\frac{\pi}{2},t\right) -\partial_x v\left(\frac{\pi}{2},t\right) =\int\limits_{-\pi/2}^{\pi/2}\tilde{R}(x,t)\, dx\,,
\end{eqnarray}
where we have denoted $\tilde{R}(x,t)\equiv\tilde{R}(U(x),x,t)$.
%\begin{equation}\label{LIN31}
%\tilde{R}(x,t)\equiv\tilde{R}(U(x),x,t).
%\end{equation}
At the initial time $t=t_0$ we have
\begin{equation}\label{t0linearized}
\int_{-\pi/2}^{\pi/2} v(x,t_0)\, dx=0\,.
\end{equation}
In the steady-state regime, that we are interested in,  the results will depend neither on the initial condition $v(x,t_0)$, nor on $t_0$ itself. Therefore, we will set $v(x,t_0)=0$ and ultimately send $t_0$ to $-\infty$.

The two conditions in Eq.~(\ref{LIN40}) can be rewritten as
\begin{eqnarray}
\label{LIN60}
&&v\left(-\frac{\pi}{2},t\right)=v\left(\frac{\pi}{2},t\right)\,,\\
\label{LIN70}
&&\delta\ell(t)=2v\left(\frac{\pi}{2},t\right)\,.
\end{eqnarray}
Equation~(\ref{LIN60}) allows us to formally extend $v$ and $\tilde R$ as functions of $x$ periodically to the whole $x$-axis. As follows from Eq.~(\ref{LIN50}), the derivative $\partial_xv$ experiences jumps at the points $x=\pi/2+m\pi$, $m=0,\pm1\,, \dots$. Each jump is equal to
\begin{equation}\label{LIN80}
\partial_x v\left(\frac{\pi}{2}+0,t\right) -\partial_x v\left(\frac{\pi}{2}-0,t\right) =\int\limits_{-\pi/2}^{\pi/2}\tilde{R}(x,t) \, dx\, .
\end{equation}
We can explicitly account for these jumps by modifying the source term in Eq.~(\ref{LIN30}):
\begin{equation}\label{LIN90}
\partial_t v(x,t)- \partial_x^2 v(x,t) -v(x,t) = R(x,t)\,,
\end{equation}
where the new source term is $R=R_b+R_d$,
\begin{eqnarray}
  R_i(x,t) &=&\! \tilde{R}_i(x,t) \nonumber \\
           &-&\!\! \left(\int\limits_{-\pi/2}^{\pi/2}\tilde{R}_i(x,t)\, dx\!\right) \!\! \sum_{m \in \mathbb{Z}}\delta\left(x-\frac{\pi}{2}-m\pi\right) \label{LIN110}
\end{eqnarray}
and $i=b, d$.  For this $R$ (and also separately for $R_b$ and $R_d$) the condition
\begin{equation}\label{LIN120}
\int\limits_{-\pi/2+\Delta}^{\pi/2+\Delta}R(x,t)\, dx=0
\end{equation}
is obeyed automatically for all $\Delta$. To avoid  uncertainty with the positions of the delta-function,
we can shift the interval of interest $\left[-\pi/2 , \pi/2 \right]$ by an infinitely small $\Delta$.
Equation~(\ref{LIN50}) now reads:
\begin{equation}\label{LIN130}
\partial_x v\left(-\frac{\pi}{2}+0,t\right) -\partial_x v\left(\frac{\pi}{2}+0,t\right) =0\,.
\end{equation}
The linearized problem is now completely defined by Eq.~(\ref{LIN90}), the conditions~(\ref{LIN60}), (\ref{LIN70}) and~(\ref{LIN130}), and a zero initial condition $v(x,t_0)=0$.

\subsection{Expanding over eigenfunctions}
\label{eigSubSec}
We can solve the linear equation~(\ref{LIN90}) by expanding the solution over the eigenfunctions $V_n(x) e^{-\lambda_n t}$ of the linear operator on the left hand side of Eq.~(\ref{LIN90}), subject to periodic boundary conditions~(\ref{LIN60}) and~(\ref{LIN130}). The complete set of eigenfunctions and eigenvalues is the following:
\begin{equation}\label{LIN140}
\left\{1, \cos(2nx),\, \sin(2nx) \right\}\,,\quad\quad
\lambda_n=4n^2-1\,,
\end{equation}
where $n=1,2,3,\dots $. Let us introduce the Green's function $G(x,x';t,t')$  of the problem, that is the solution of Eq.~(\ref{LIN90}) with the source term $R=\delta(x-x')\delta(t-t')$, subject to the boundary conditions ~(\ref{LIN60}) and~(\ref{LIN130})  and a zero initial condition at $t=t'$. The Green's function has the form
\begin{eqnarray}
  G(x,x';t,t')&=& \left[\frac{e^{t-t'}}{\pi}+ \frac{2}{\pi}\sum\limits_{n=1}^\infty  \cos 2n(x-x')\,  e^{-\lambda_n (t-t')}\right] \nonumber \\
  &\times& \theta(t-t')\,,
\end{eqnarray}
where $\theta(t-t')$ is the step function.  Using the Green's function, we can write an explicit expression for $v(x,t)$ and, by virtue of Eq.~(\ref{LIN70}), for $\delta\ell(t)$. Sending $t_0$ to $-\infty$, we arrive at the following results:
\begin{equation}\label{LIN160}
v(x,t)\!=\!\frac{2}{\pi}\sum\limits_{n=1}^\infty \int\limits_0^\infty e^{-\lambda_n t'} \!\!\!\!\!\!\!\! \int\limits_{-\pi/2+0}^{\pi/2+0} \!\!\!\!\!\! R(x',t-t')\, \cos 2n(x-x')\, dx' dt',
\end{equation}
\begin{equation}\label{LIN170}
\delta\ell(t)\!=\!\frac{4}{\pi}\sum\limits_{n=1}^\infty (-1)^n    \int\limits_0^\infty e^{-\lambda_n t'} \!\!\!\!\!\!\!\!\int\limits_{-\pi/2+0}^{\pi/2+0}\!\!\!\!\!\!
R(x,t-t') \cos 2nx\, dx\, dt' .
\end{equation}
We will use these expressions in the subsequent sections \ref{comSec} and \ref{lSec} to analyze the fluctuations of the swarm center of mass (COM) and of the swarm radius, respectively.

\begin{figure}
\includegraphics[width=6.5cm]{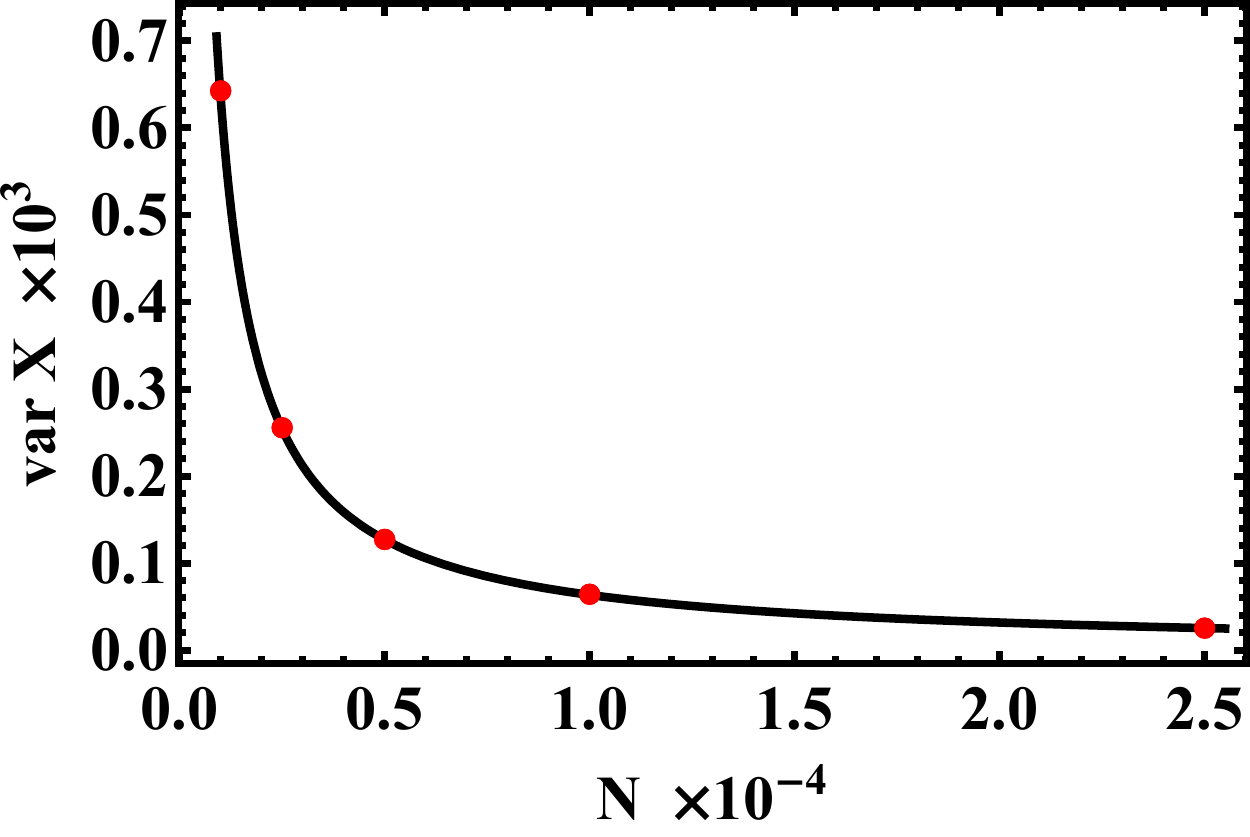}
\includegraphics[width=6.5cm]{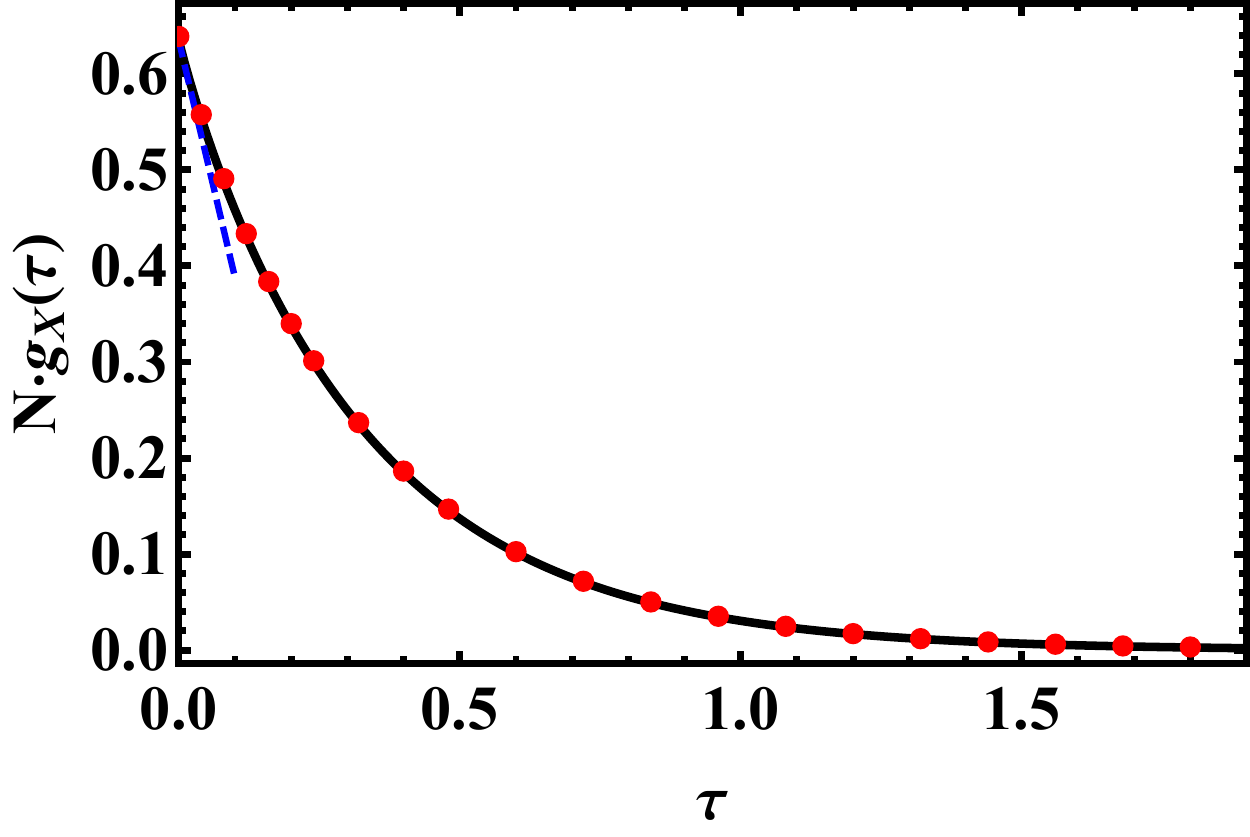}
\caption{Fluctuations of the COM $X(t)$. Top panel: the variance of $X(t)$ versus $N$ from Monte-Carlo simulations (symbols) and from theoretical prediction (\ref{COM60}) (solid line). Bottom panel: the  covariance $g_X (\tau)$ vs. $\tau$ at $N=2500$ from simulations (symbols) and from Eq.~(\ref{COM100}) (solid line). The dashed line is the small-$\tau$ asymptotic
$g_X (\tau)=  0.6346... - 2.4674... \,\tau$.}
\label{COMfig}
\end{figure}

\section{Fluctuations of the center of mass}
\label{comSec}
In the Langevin description, the coordinate of the COM of the system is
\begin{equation}\label{COM10}
X(t) = \int_{-\ell(t)}^{\ell(t)} x u(x,t) dx = \int_{-\pi/2}^{\pi/2} x v(x,t) \, dx\,,
\end{equation}
where, to the leading order, we can use our result (\ref{LIN160}) for $v(x,t)$. Let us evaluate the COM's covariance in the steady state,
\begin{equation}\label{COM20}
g_X(\tau) \equiv \textrm{cov(} X(t),X(t+\tau) \textrm{)}=\left<X(0)X(\tau)\right>\,.
\end{equation}
Combining Eqs.~(\ref{LIN160}) and~(\ref{COM10}), we obtain
\begin{eqnarray}
  X(t) &=& \sum_{n=1}^\infty \frac{(-1)^{n+1}}{n} f_n(t)\,,\nonumber  \\
 f_n(t) \!&=&\! \int\limits_0^\infty e^{-\lambda_n t'} \!\!\!\!\!\! \int\limits_{-\pi/2+0}^{\pi/2+0} \!\!\!\! R(x,t-t')\, \sin(2nx)\, dx\, dt'. \label{COM30}
\end{eqnarray}
Plugging $X(t)$ into Eq.~(\ref{COM20}) and performing the averaging over the noises $\eta$ and $\chi$,
we obtain after some algebra
\begin{equation}\label{COM40}
g_X(\tau)=\frac{1}{N}\sum_{n,\, m\geq 1} s^{(X)}_{nm}\,,
\end{equation}
where
\begin{equation}\label{COM50}
s^{(X)}_{nm}\!=\! \frac{-16(n^2+m^2)\,\exp\left[ -\left(4n^2-1\right) |\tau| \right]}{(4(n+m)^2-1)(4(n-m)^2-1)(2n^2+2m^2-1)}\,.
\end{equation}
Fortunately, the infinite sum over $m$ can be evaluated exactly using ``Mathematica". This leaves us with a single infinite sum:
\begin{equation}\label{COM100}
g_X(\tau)=\frac{1}{N} \sum_{n=1}^{\infty} r_n^{(X)}\,\exp\left[ -\left(4n^2-1\right) |\tau| \right]\,,
\end{equation}
where
\begin{widetext}
\begin{equation}\label{COM110}
r_n^{(X)}=\frac{8 \left(64 n^6-32 n^4+n^2\right)-2 \pi  \left(4 n^2-1\right)^2
   \sqrt{4 n^2-2}\,\coth \left(\pi
   \sqrt{n^2-\frac{1}{2}}\right)}{\left(4 n^2-1\right)^2 \left(128
   n^6-128 n^4+34 n^2-1\right)}\,.
\end{equation}
\end{widetext}
As $n\to \infty$, the coefficient $r_n^{(X)}$ behaves as $1/n^4$. Therefore, the infinite sum in Eq.~(\ref{COM100}) converges for any $\tau$ including $\tau=0$.  At long times, $|\tau|\gg 1$, $g_X(\tau)$ decays exponentially with the decay rate determined by the ground state eigenvalue $\lambda_1=3$:
\begin{equation}\label{COM120}
g_X (|\tau| \gg 1) \simeq \frac{1}{N}\, r_1^{(X)} e^{-3|\tau|}\,,
\end{equation}
where
\begin{equation}\label{COM130}
 r_1^{(X)}=\frac{264-18 \sqrt{2} \pi  \coth \left(\frac{\pi
   }{\sqrt{2}}\right)}{297}= 0.61321\dots \,.
\end{equation}
In its turn, the quantity $g_X(\tau=0)$ yields the variance of the COM fluctuations:
\begin{equation}\label{COM55}
\left<X^2\right>=g_X(\tau=0)=\frac{1}{N} \sum_{n=1}^{\infty} r_n^{(X)}\,.
\end{equation}
Evaluating the infinite sum over $n$
numerically, we obtain
\begin{equation}\label{COM60}
\text{var}\, X(t) = \left<X^2\right>=\frac{0.6346 \dots}{N}\,.
\end{equation}
Figure~\ref{COMfig} (top panel) compares this theoretical prediction  with the variance of the COM, measured in Monte-Carlo simulations at different $N$ (see the Appendix for a brief description of the simulation algorithm). Figure ~\ref{COMfig} (bottom panel) compares the theoretically predicted covariance $g_X (\tau)$, Eq.~(\ref{COM40}), with the covariance measured in the simulations at $N=2500$ and different $\tau$. An excellent agreement is observed in both cases.

The $1/N$ scaling of the variance of the COM is to be expected from the law of large numbers. It also reflects the fact that large-scale hydrodynamic modes (with wavelengths comparable to the swarm size) dominate the contribution to the variance. We will encounter quite a different situation in the next section, which deals with fluctuations of the swarm radius $\ell(t)$.

\section{Fluctuations of the swarm radius}
\label{lSec}
Averaging Eq.~(\ref{LIN170}) over the noise, we obtain $\left<\delta\ell(t)\right>=0$. Therefore, at least to the accuracy of the linear theory in $1/\sqrt{N}$, we have $\left<\ell(t)\right>=\pi/2$.
%\begin{equation}\label{L20}
%\left<\ell(t)\right>=\frac{\pi}{2}\,.
%\end{equation}
The covariance of $\ell(t)$ is given by
\begin{equation}\label{L30}
g_\ell(\tau) = \left<\ell(0)\ell(\tau)\right>-\left<\ell\right>^2 = \left<\delta\ell(0)\delta\ell(\tau)\right>\,.
\end{equation}
It is convenient to calculate the contributions to $g_\ell(\tau)$ from the branching noise and from the noise of Brownian motion separately. Therefore we define $\delta\ell=\delta\ell_b+ \delta\ell_d$, where
\begin{equation}\label{L40}
\delta\ell_i(t)\!=\!\frac{4}{\pi}\sum\limits_{n=1}^\infty (-1)^n\!\!    \int\limits_0^\infty e^{-\lambda_n t'} \!\!\!\!\!\!\!\!\int\limits_{-\pi/2+0}^{\pi/2+0}\!\!\!\!\!\!
R_i(x,t-t') \cos 2nx\, dx \, dt'
\end{equation}
and $i=b, d$. We can write
\begin{eqnarray}
  g_\ell(\tau) &=& g_\ell^{(b)}(\tau)+g_\ell^{(d)}(\tau)\nonumber \\
  &=& \left<\delta\ell_b(0)\,\delta\ell_b(\tau)\right>+\left<\delta\ell_d(0)\,\delta\ell_d(\tau)\right> \, .
\end{eqnarray}
Then, using Eq.~(\ref{L40}) and performing the averaging, we obtain
\begin{equation}\label{L60}
g_\ell^{(i)}(\tau)=\frac{8}{\pi^2 N} \sum_{n,\, m\geq 1} \frac{\exp\left[-\left(4n^2-1\right) |\tau| \right]}{4(n^2+m^2)-2}  s^{(i)}_{nm}\,,
\end{equation}
where
\begin{eqnarray}
  s^{(b)}_{nm} &=&  -\frac{1}{4(n+m)^2-1}-\frac{1}{4(n-m)^2-1} \nonumber\\
 &+& \frac{2}{4n^2-1}+\frac{2}{4m^2-1}+2 \label{L80}
\end{eqnarray}
and
\begin{equation}\label{L70}
s^{(d)}_{nm}= 8nm \left[ \frac{1}{4(n+m)^2-1}-\frac{1}{4(n-m)^2-1} \right]\,.
\end{equation}
Again, the infinite sums over $m$ can be evaluated exactly with ``Mathematica",  leading to the expression
\begin{equation}\label{L90}
g_\ell^{(i)}(\tau)=\frac{64}{\pi N} \sum_{n=1}^\infty r_n^{(\ell)} \exp\left[ -\left(4n^2-1\right) |\tau| \right]\,,
\end{equation}
where
\begin{equation}\label{L100}
r_n^{(\ell)}=\frac{n^2\sqrt{4n^2-2}\coth(\pi \sqrt{n^2-1/2})}{64n^4-32n^2+1}
\end{equation}
for both $i=b$ and $i=d$. Somewhat surprisingly, the branching noise and the noise of the Brownian motion give identical contributions to the covariance of $\ell(t)$.
Therefore,
\begin{equation}\label{L95}
g_\ell(\tau)=\frac{128}{\pi N} \sum_{n=1}^\infty r_n^{(\ell)} \exp\left[ -\left(4n^2-1\right) |\tau| \right]\,.
\end{equation}

For $\tau\neq 0$ the infinite sum over $n$ in Eq.~(\ref{L95}) converges. In particular, the long-time decay of $g_\ell(\tau)$ with $\tau$ is again purely exponential, $\sim e^{-3|\tau|}$, with the same ground state eigenvalue $\lambda_1=3$ as for the COM.

But what happens at $\tau=0$? The large-$n$ asymptotic of $r_n^{(\ell)}$ from Eq.~(\ref{L100}) is
\begin{equation}\label{L110}
r_n^{(\ell)} \simeq \frac{1}{32 n}\,.
\end{equation}
Therefore, at $\tau=0$ the sum over $n$ in Eq.~(\ref{L95}) diverges logarithmically at $n\to \infty$, implying an infinite variance. On the other hand, our Monte Carlo simulations (see below) clearly show a finite variance of $\ell$. How to resolve this paradox?

\begin{figure}[t]
%\centering
\includegraphics[width=6.5cm]{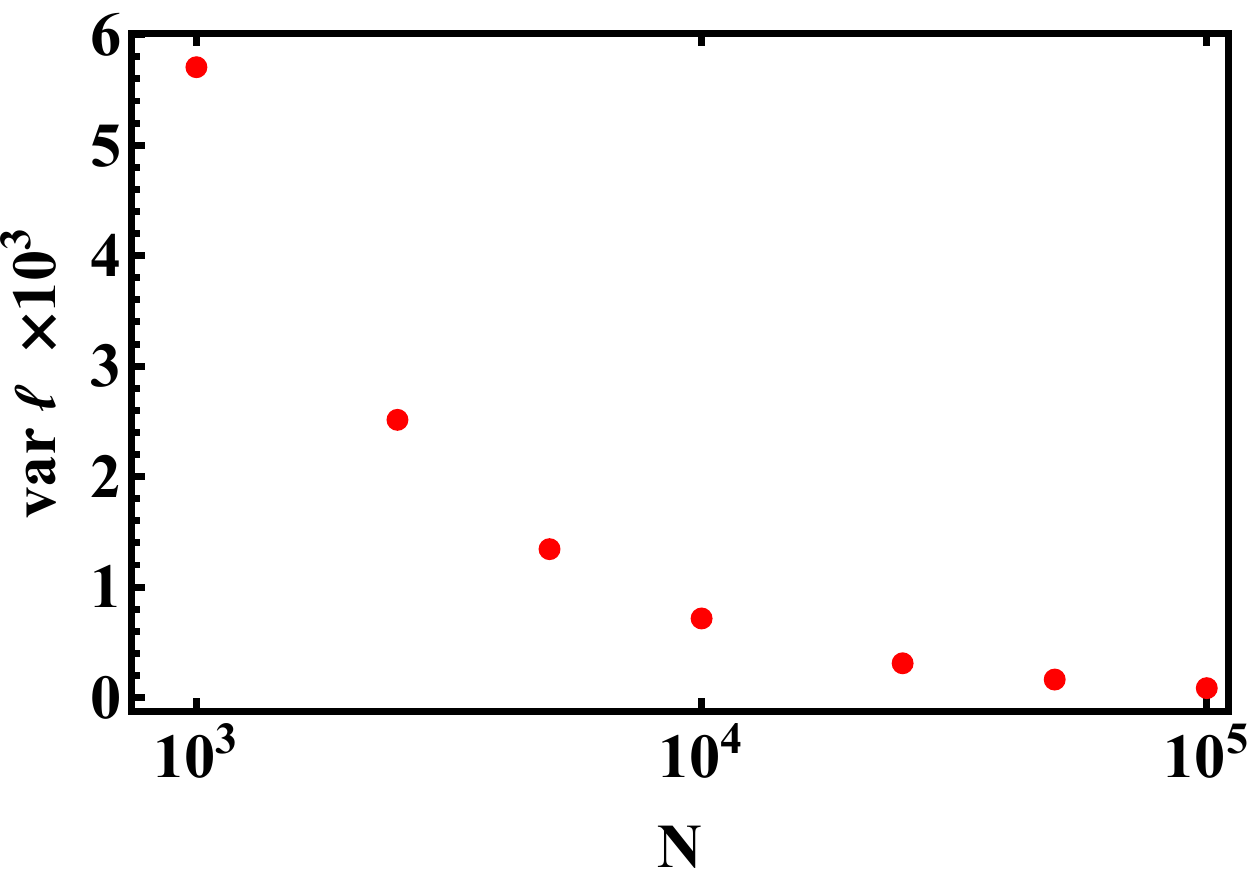}
\includegraphics[width=6.5cm]{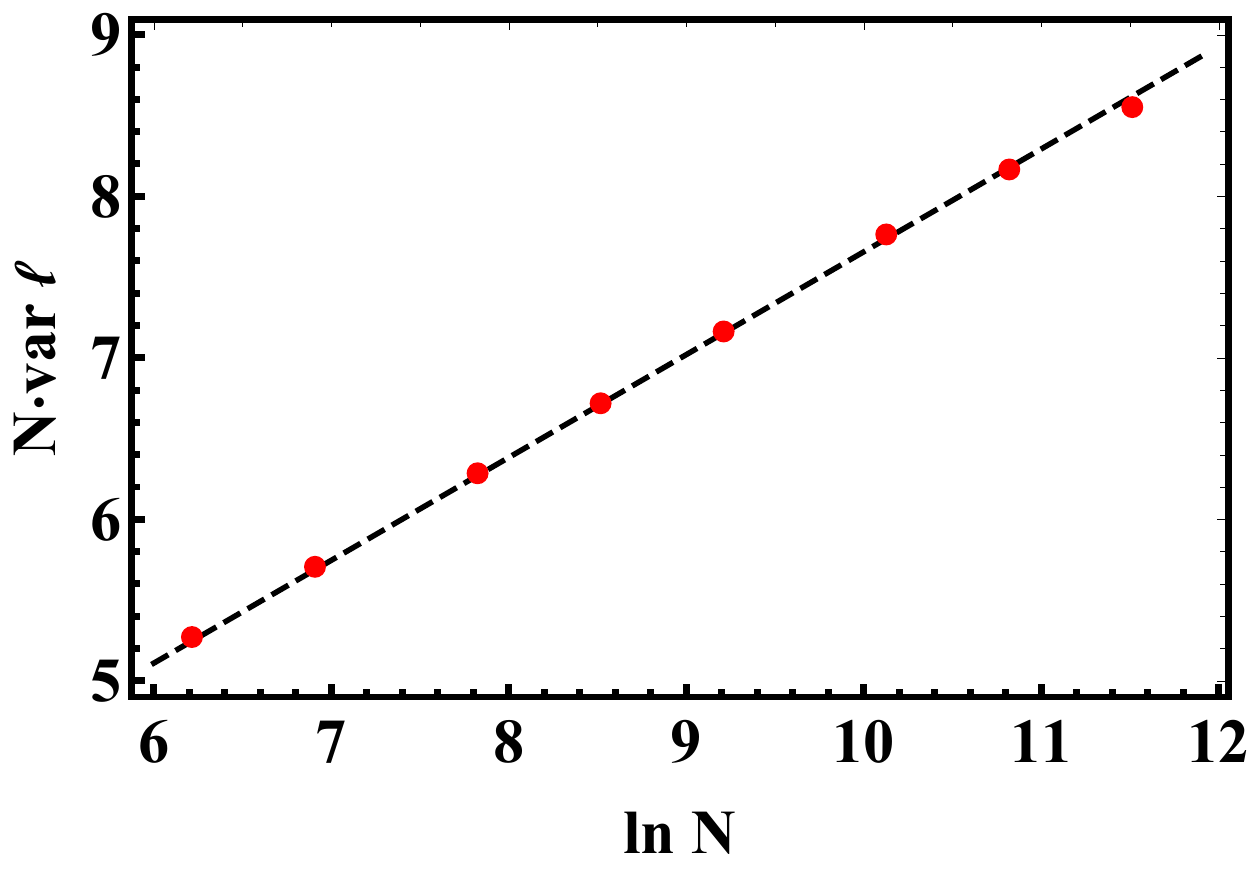}
\caption{Fluctuations of the swarm radius $\ell(t)$. Top panel: the variance of $\ell(t)$ versus $N$ as measured in Monte Carlo simulations. Bottom panel: $N$ times the variance of $\ell(t)$ vs. $\ln N$. Symbols indicate simulation results. The dashed line shows the function $(2/\pi) \ln N+a$, where $a\simeq 1.29  \pm 0.03$, see the text after Eq.~(\ref{L140}).}
\label{LvarFig}
\end{figure}

Let us return to the covariance $g_{\ell}(\tau)$ and recall that the Langevin description,  which we adopted here as an approximation to the microscopic model,  is applicable only at ``macroscopic" times $\tau \gg 1/N$.  Since the divergence of $g_{\ell}(\tau)$ as $\tau\to 0$ is only logarithmical, we can introduce a cutoff at $\tau \sim 1/N$ in the hope to obtain
correct result for the variance of $\ell(t)$ with logarithmic accuracy. Therefore, we proceed in the following way.
For $1/N \ll |\tau|\ll 1$ we use the large-$n$ asymptotic (\ref{L110}) of $r_n^{(\ell)}$
and replace the summation over $n$ in Eq.~(\ref{L95}) by integration. This gives
\begin{equation}\label{L120}
g_{\ell}(\tau) \simeq \frac{2}{N} \frac{e^{|\tau|}}{\pi} \Gamma(0,4|\tau|)\,,
\end{equation}
where $\Gamma(0,4|\tau|)$ is the incomplete Gamma function. In fact, we must use the  $|\tau| \ll 1$ asymptotic  of Eq.~(\ref{L120}), which leads us to
\begin{equation}\label{L130}
g_\ell(\tau)\simeq \frac{2}{\pi N} \ln \frac{1}{|\tau|}\,,\quad \frac{1}{N}\ll |\tau|\ll 1\,.
\end{equation}
The logarithmic growth of correlations of $\ell(t)$ with a decrease of the time difference $\tau$ is a noticeable phenomenon. In the frequency space it corresponds  to a $1/f$ noise. The $1/f$ noise has been observed in a plethora of physical and biological systems \cite{1fnoise}, and its appearance in the fluctuations of $\ell(t)$ is unexpected.

\begin{figure}[t]
%\centering
\includegraphics[width=6.5cm]{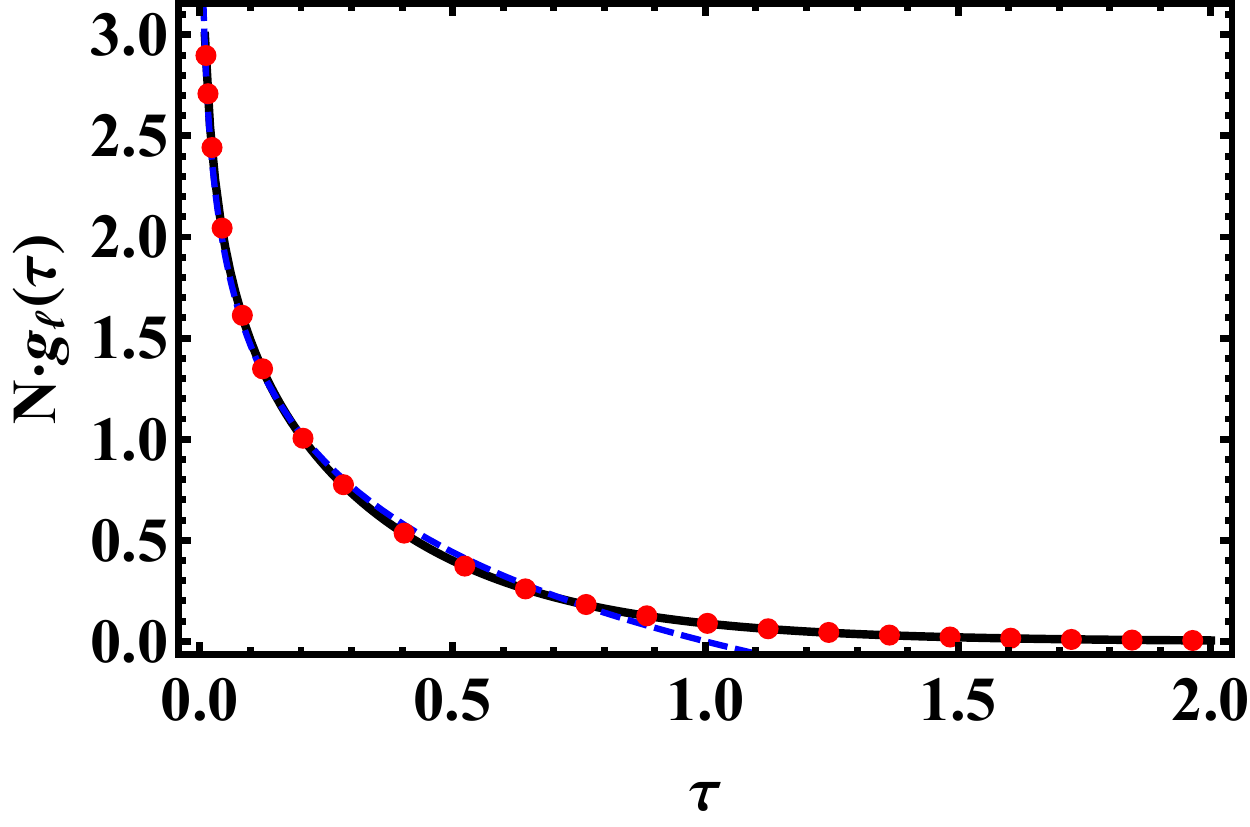}
\caption{Fluctuations of the swarm radius $\ell(t)$. Symbols: the covariance $g_{\ell}(\tau)$ measured in the Monte Carlo simulations for  $N\!=\!2500$. Solid line: theoretical prediction (\ref{L95}). Dashed line: the asymptotic~(\ref{L130}).}
\label{LcovFig}
\end{figure}

Now we can evaluate the variance of $\ell(t)$ by introducing a cutoff in Eq.~(\ref{L130}) at $\tau=1/N$. We obtain, with logarithmic accuracy,
\begin{equation}\label{L140}
\textrm{var }\ell(t) \simeq \frac{2}{\pi} \frac{\ln N}{N}\,.
\end{equation}
This result is markedly different from Eq.~(\ref{COM60}) by the presence of the large logarithmic factor $\ln N$, and this result is fully supported by our Monte-Carlo simulations of the microscopic model. Figures~\ref{LvarFig} and~\ref{LcovFig} show the simulation results for the fluctuations of $\ell(t)$, and compare them with theoretical predictions  (\ref{L95}), (\ref{L130}) and (\ref{L140}). As one can see from the bottom panel of Fig.~\ref{LvarFig}, the slope of the straight line, predicted by Eq.~(\ref{L140}), agrees very well with the simulations. To eliminate the offset, we introduced a single adjustable offset parameter $a$. It corresponds to a numerical factor under the logarithm which is beyond the logarithmic accuracy of Eq.~(\ref{L140}).

\section{Summary and Discussion}
\label{sumSec}

We studied analytically and numerically fluctuations of a swarm of Brownian bees in the limit of large but finite $N$. We focused on two fluctuating quantities: the COM of the swarm $X(t)$ and the swarm radius $\ell(t)$. We employed a first-principles Langevin equation for this system, linearized it around the deterministic steady state solution, and calculated the two-time covariances of $X(t)$ and $\ell(t)$. As we found, the variance of $X(t)$ behaves ``normally": it scales as $1/N$ as to be expected from the law of large numbers. But the variance of the swarm radius $\ell(t)$ exhibits an anomalous scaling $(1/N) \ln N$. This anomaly is caused by significant contributions to the variance from all spatial scales: from scales comparable to the swarm size down to the scale $\sim 1/\sqrt{N}$ of a narrow region near the edges of the swarm where only a few  bees are present.
We have also shown that, at high but not too high frequencies, the power spectrum of the fluctuations of $\ell(t)$ exhibits the $1/f$ noise. Our Monte-Carlo simulations of the microscopic system are in good agreement with theory.

\begin{figure}[t]
%\centering
\includegraphics[width=6.5cm]{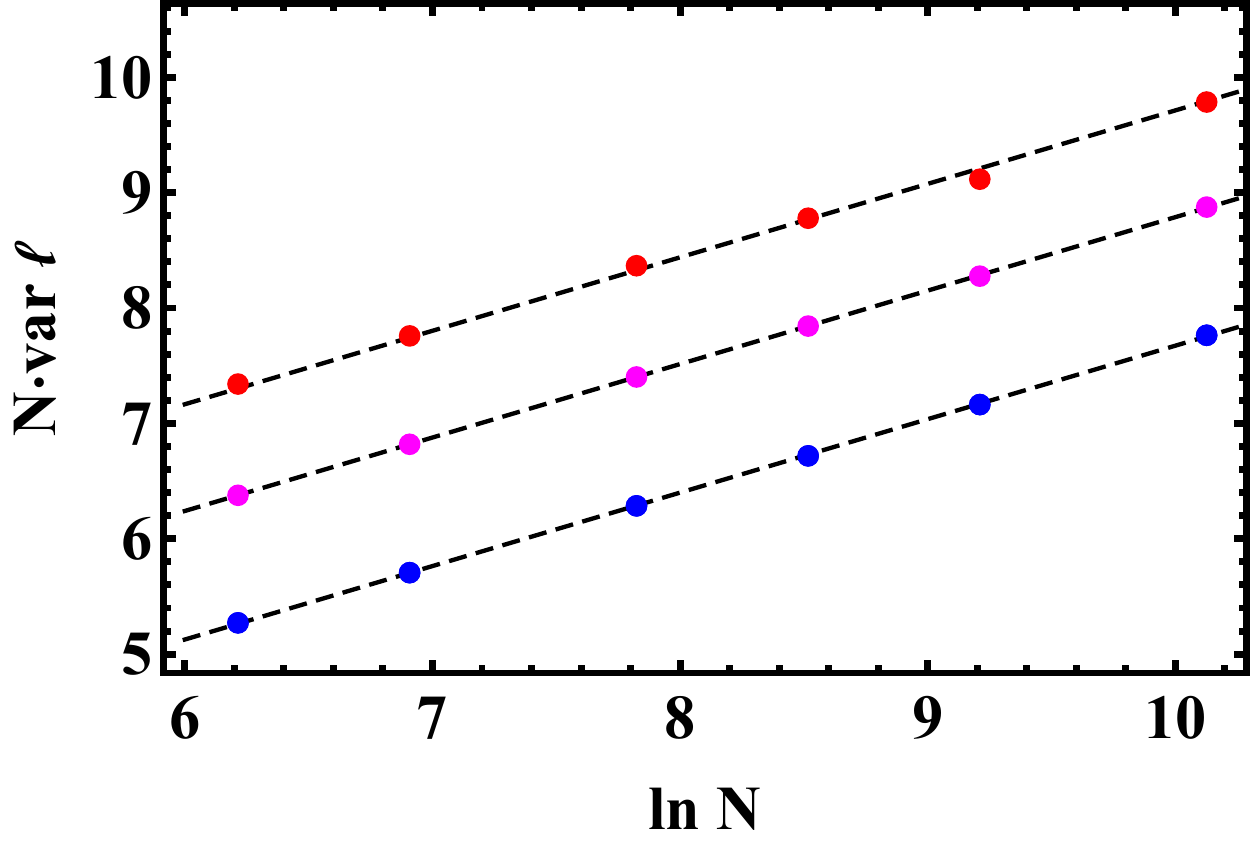}
\caption{The variance of the swarm radius $\ell(t)$ for $d=1,2$ and $3$ as measured in the Monte Carlo simulations. Shown is $N$ times the variance of $\ell(t)$ vs. $\ln N$.  The symbols indicate simulation results for $d=1$, $2$ and $3$ (from bottom to top, respectively). The three dashed lines show the function $(2/\pi) \ln N+a_d$ with the same slope $2/\pi$ and $d$-dependent offsets: $a_1=1.29  \pm 0.03$, $a_2=2.42  \pm 0.01$, and $a_3=3.35  \pm 0.05$.}
\label{Lvar123Fig}
\end{figure}

A generalization to higher dimensions is especially interesting for the fluctuations of the swarm radius $\ell(t)$. One should expect that a small fraction of particles near the (circular or spherical) swarm boundary give a significant contribution to the variance of $\ell(t)$ in higher dimensions as well. Furthermore,  in all dimensions, the average steady-state particle flux to the boundary is the same (and equal to $N$). Therefore, the  deterministic steady-state particle density $u=U_d(\mathbf{x})$  behaves in a universal way near the swarm boundary. Does this universality lead to a similar behavior of the variance of $\ell(t)$ vs. $N$ in different dimensions? Figure \ref{Lvar123Fig} depicts the results of our Monte Carlo simulations for $d=2$ and $3$, where we measured the variance of $\ell(t)$ at different $N$. For comparison, also shown is the variance of $\ell(t)$ for $d=1$. Amazingly, as one can see from the figure, the slope $2/\pi$ of the straight line
$N \times \text{var} \,\ell(t)$ vs. $\ln N$, predicted by Eq.~(\ref{L140}) for $d=1$, holds for $d=2$ and $3$ as well. This remarkable coincidence certainly deserves a closer look at via an extension of our analytical calculations to higher dimensions.

It is instructive to compare the Brownian bees model with stochastic models, describing ``pulled" reaction fronts propagating into an unstable empty state. The best known example of such models is the stochastic Fisher-Kolmogorov-Petrovsky-Piskunov (F-KPP) equation. The velocity of a stochastic F-KPP front fluctuates around its mean value, and the variance of these fluctuations scales as $1/\ln^3 N$, where $N\gg 1$ is the typical number of particles in the front region \cite{Derridaetal}. This F-KPP anomaly is much stronger than the one in the Brownian bees model. This is because the F-KPP front velocity fluctuations are \emph{solely} determined by a small fraction of particles located at the leading edge of the front \cite{Derridaetal}.

In conclusion, we should remind the reader that the Langevin equation provides an accurate description only of typical, small fluctuations of reacting and diffusing systems of particles. Therefore, our present work is limited to typical  fluctuations  of the Brownian bees. It would be very interesting, and challenging, to also study \emph{large deviations} of $X(t)$ and $\ell(t)$, which correspond to the distribution tails of these quantities. Large deviations of \emph{persistent} fluctuations of $\ell(t)$ have been recently addressed in Ref. \cite{MS2021} by employing the optimal fluctuation method which bypasses the Langevin description.

\vspace{0.5 cm}
\section*{Acknowledgments} We are grateful to N. R.  Smith for useful discussions. The research of M.S. and B.M. was supported by the Israel Science Foundation (grant No. 1499/20). The research of P.S. was supported by the project ``High Field Initiative" (CZ.02.1.01/0.0/0.0/15\_003/0000449) of the European Regional Development Fund.

\appendix

\section{Monte-Carlo simulations}
\label{simulations}

We performed continuous-time Monte-Carlo simulations (see \textit{e.g.} Ref. \cite{MCbook}) of the microscopic model. The small time intervals $\Delta t$ between two consecutive binary branching events were drawn from the exponential distribution with rate parameter $N$. During these time intervals $\Delta t$ the particle positions were advanced according to the Brownian motion, given by the normal distribution $N(x_i,2\Delta t)$, where $x_i$ is the particle's initial position, and $i=1,2,\dots, N$.

Next, one particle is chosen randomly, from the uniform distribution, to be the particle which is about to branch. The particle farthest from the origin is then relocated to  the position of the chosen particle. This  is equivalent to the removal of the farthest particle because of the birth of a ``new" particle. This process can be thought of as relabeling the particles.

The measurements  were performed, after the system reached a steady state, at fixed time steps $\delta t=1/N$.  The simulation time was chosen to be sufficiently large, so that the precision of all measurements is at least within $1 \%$ with $95\%$ confidence.

\end{document}